\documentclass{article}

\PassOptionsToPackage{numbers}{natbib}

\usepackage[final]{nips_2018}

\usepackage[utf8]{inputenc} 
\usepackage[T1]{fontenc}    
\usepackage{hyperref}       
\usepackage{url}            
\usepackage{booktabs}       
\usepackage{amsfonts}       
\usepackage{nicefrac}       
\usepackage{microtype}      
\usepackage{amsmath}  
\usepackage{graphicx} 
\usepackage{multirow} 
\usepackage{siunitx}  
\usepackage{setspace} 
\usepackage{fancyhdr} 
\usepackage{pdfpages}

\usepackage[ruled,linesnumbered]{algorithm2e}
\usepackage[bottom]{footmisc}

\title{On the human evaluation of audio adversarial examples}

\author{
  Jon Vadillo and Roberto Santana
     \\
  Department of Computer Science and Artificial Intelligence \\
  Faculty of Informatics\\
  University of the Basque Country (UPV/EHU)\\
  \texttt{jvadillo005@ikasle.ehu.eus , roberto.santana@ehu.eus} \\
}

\providecommand{\keywords}[1]{\textbf{\textit{Keywords: }} #1}

\usepackage{tikz}

\begin{document}
\cleardoublepage

\maketitle


\begin{abstract}
Human-machine interaction is increasingly dependent on speech communication. Machine Learning models are usually applied to interpret human speech commands. However, these models can be fooled by adversarial examples, which are inputs intentionally perturbed to produce a wrong prediction without being noticed. While much research has been focused on developing new techniques to generate adversarial perturbations, less attention has been given to aspects that determine whether and how the perturbations are noticed by humans. This question is relevant since high fooling rates of proposed adversarial perturbation strategies are only valuable if the perturbations are not detectable. In this paper we investigate to which extent the distortion metrics proposed in the literature for audio adversarial examples, and which are commonly applied to evaluate the effectiveness of methods for generating these attacks,  are a reliable measure of the human perception of the perturbations. Using an analytical framework, and an experiment in which 18 subjects evaluate audio adversarial examples, we demonstrate that the metrics employed by convention are not a reliable measure of the perceptual similarity of adversarial examples in the audio domain.
\end{abstract}

\keywords{Adversarial Examples, Deep Neural Networks, Speech Command Classification, Speech Recognition}

\section{Introduction}
\label{section::introduction}

Human-computer interaction increasingly relies on Machine Learning (ML) models such as Deep Neural Networks (DNNs) trained from,  usually large,  datasets \cite{Fang_et_al:2018, Gao_et_al:2019, Hassan_et_al:2018, Nunez_et_al:2018}.  The ubiquitous applications of DNNs in security-critical tasks, such as face identity recognition \cite{sun2014deep, parkhi2015deep}, speaker verification \cite{heigold2016end, snyder2017deep}, voice controlled systems \cite{feng2017continuous,boles2017voice,gong2018overview} or signal forensics \cite{bayar2018constrained, bayar2017generic, zeng2017deep, athulya2017mitigating} require a high reliability on these computational models. However, it has been demonstrated that such models can be fooled by perturbing an input sample with malicious and quasi-imperceptible perturbations. These attacks are known in the literature as adversarial examples \cite{Szegedy:42503,goodfellow2014explaining}. Due to the fact that these attacks are designed to be hardly detectable, they suppose a serious concern regarding the reliable application of DNNs in adversarial scenarios.

The study of adversarial examples has focused primarily on the image domain and computer vision tasks \cite{akhtar2018threat}, whereas domains such as text or audio have received much less attention. In fact, such domains imply additional challenges and difficulties. One of the evident differences between domains is the way in which the information is represented, and, therefore, the way in which adversarial perturbations are measured, bounded and perceived by human subjects. 
In the image domain, $L_p$ norms are mainly used as a basis to measure the distortion between the original signal and the adversarial example. However, recent works have pointed out that such metrics do not always properly represent the perceptual distortion introduced by adversarial perturbations \cite{fezza2019perceptual, jordan2019quantifying,dukler2019wasserstein}.
Although in some works $L_p$ norms are also used during the generation of adversarial examples to limit the amount of perturbation \cite{alzantot2018did, gong2017crafting}, in the audio domain more representative metrics are usually employed for acoustic signals, such as signal-to-noise ratio (SNR) \cite{yakura2018robust,du2019sirenattack} or Sound Pressure Level (SPL) \cite{zhang2017dolphinattack,roy2017backdoor,abdoli2019universal}. These metrics are computed in decibels (dB), which is a standard scale employed for acoustic signals. However, even for such metrics, measuring the perceptual distortion of the attacks is not straightforward, as other characteristics have a high influence, such as time-frequency properties.
In text problems the difficulty of characterizing the perceptual distortion is even greater, due to the fact that every change is inevitably noticeable, and therefore, the aim is to produce semantically and syntactically similar adversarial examples \cite{alzantot-etal-2018-generating}.

In this paper we focus on the human evaluation of adversarial examples in the audio domain. A more comprehensive approach to evaluating adversarial distortions can serve to better understand the risks of adversarial attacks in the audio domain. For instance, the development of adversarial defenses or secure human machine interaction systems can focus on the more effective, unnoticeable, attacks. Therefore, the goal of this study is to perform an analysis of the human perception of audio adversarial perturbations according to different factors. Based on these results, we will also study to which extent the similarity-metrics employed in the literature are suitable to model such subjective criterion.  

The remainder of the paper is organized as follows: In the following section we introduce the main concepts related to adversarial examples and review previous approaches to evaluate the distortion produced by adversarial perturbation in the audio domain. This section also highlights a number of research questions related to the evaluation of audio distortion that have not been previously addressed.
Section~\ref{section:experimental_setup} describes the selected task, target models and dataset, as well as the particular method employed for generating adversarial perturbation in the audio domain. Section~\ref{section:evaluation_metrics} presents a preliminary evaluation of the adversarial perturbation according to the metrics proposed in the literature.
In Section~\ref{section:human_experiment_distortion}, we present the design of an experiment to find answers to some of the issues involved in the perceptual evaluation of the perturbations. The results of the experiment in which $18$ human subjects evaluate different aspects of the adversarial perturbations are also presented and discussed. Section~\ref{section:conclusions} concludes the paper and identifies lines for future research.

\section{Related work} 
\label{section:relatedwork}

\begin{figure}[]
\centering
\includegraphics[scale=0.3]{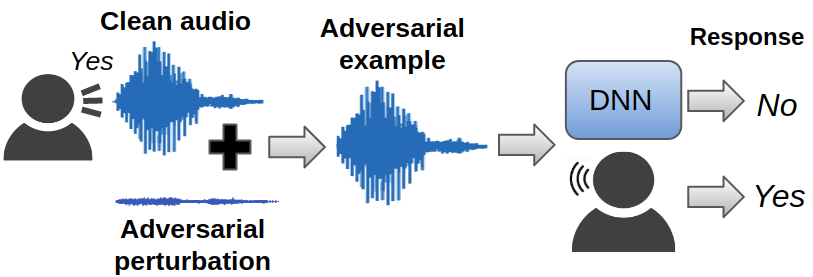}
\caption{Illustration of an adversarial attack, in which an adversarial perturbation is added to a clean audio waveform, forming an \textit{adversarial example} which is misclassified by a target DNN model, while not altering the human perception of the audio.}
\label{fig:attack_scheme}
\end{figure}

The existence of adversarial examples which are able to fool DNNs have been reported for many different audio related tasks, such as automatic speech recognition \cite{alzantot2018did, carlini2018audio, neekhara2019universal}, music content analysis \cite{kereliuk2015deep} or sound classification \cite{abdoli2019universal}.   A common adversarial attack scheme is represented in Fig. \ref{fig:attack_scheme}. Note that it is assumed that an adversary can feed the perturbed signal directly into the model. Even if this is a common assumption, some works have demonstrated that such attacks can be designed to work in the physical world \cite{yakura2018robust, carlini2016hidden, yuan2018commandersong}. 

Our work builds on previous research where adversarial perturbations for audio command classification have been introduced \cite{Vadillo_and_Santana:2019}. However, the evaluation methods and type of analysis presented in this paper are valid for other approaches conceived to generate audio adversarial examples for other ML tasks.

\subsection{Adversarial example: formal description} 
\label{section:attack_formal_description}

Let $f(x)$ be a classification model $f : \mathbb{X}\rightarrow \mathbb{Y}$, which classifies an input $x$ from the input space $\mathbb{X}$ as one of the classes represented in $\mathbb{Y}=\left\{ y_1,...,y_k\right\}$.
An adversarial example $x'$ is defined as $x'=x+v$, where $v \in \mathbb{R}^d$ represents the adversarial perturbation capable of producing a misclassification of $f$ for the (correctly classified) input $x$: $f(x)\neq f(x')$. A necessary requirement for an adversarial attack is that the perturbation should be \textit{imperceptible}, and therefore, the goal is to minimize the distortion introduced by $v$ as much as possible, according to a suitable distortion metric $\varphi(x,x')\rightarrow \mathbb{R}$

Depending on the objective of the attack, adversarial examples can be categorized in different ways. First of all, a \emph{targeted} adversarial example consists of a perturbed sample $x'=x+v$ which satisfies $f(x')=t$, where $t$ represents the target (incorrect) label that we want to be produced by the model. In contrast, an \emph{untargeted} adversarial example only requires the output label to be incorrect  $f(x')\neq f(x)$, without any additional regard about the output class assigned to $x'$.

Furthermore, depending on the scope of the adversarial perturbation $v$, we can differentiate between \emph{individual} or \emph{universal} adversarial perturbations. In the first case, the perturbation is crafted specifically to be applied to one particular input $x$. Therefore, it is not expected that the same perturbation will be able to fool the model for a different sample. In the second case, universal adversarial examples are \emph{input agnostic} perturbations able to fool the model independently of the input. In \cite{Vadillo_and_Santana:2019}, different levels of universality are proposed, depending on the number of classes for which it is expected to work. The first universality level comprises \textit{single-class} universal perturbations that are conceived to fool the target model only for inputs of one particular class. We will focus on \textit{single-class} universal perturbations, although our findings regarding the weaknesses and gaps in the evaluation of adversarial perturbations are not restricted to this universality level.

\subsection{Methods for assessing audio adversarial perturbations} \label{section:previous_works_audio}

In this section we collect the strategies employed by previous works in order to verify that audio perturbations are not detectable by humans. Even if an essential requirement for adversarial perturbations to suppose a real threat is that they must be imperceptible, a good specification of such (mainly subjective) a constraint is not straightforward, and, indeed, is not well established yet. 

Furthermore, even if the analysis is constrained to the audio domain, the understanding and definition of what can make a sample natural is very related to the ML task that is being solved by the model (e.g., it might be harder to categorize a music tune as ``unnatural'' than a spoken command). With a large variety of ML tasks related to the analysis of acoustic signals (e.g., speech recognition, music content analysis or ambient sound classification), each of them may require, therefore, a different criterion to assess the distortion of the adversarial examples according to human perception. Although a number of strategies have been proposed in these domains \cite{roy2017backdoor, kereliuk2015deep, zhang2017dolphinattack, schonherr2018adversarial, carlini2018audio}, we focus on those suitable for spoken commands. Among these strategies are:

\begin{itemize}
 \item Thresholding the perturbation amount
 \item Models of human perception and hearing system
 \item Human evaluation
\end{itemize}

\subsubsection{Thresholding the perturbation amount}
\label{section:assessing_thresholding}
The methods discussed in this section rely on limiting or measuring the perturbation amount that is added to the original input, according to a distortion metric, to ensure that the perturbations are imperceptible or \textit{quasi-imperceptible}, or that the distortion levels are below a maximum acceptable threshold.

In \cite{alzantot2018did}, the perturbation applied to spoken commands is restricted to the 8 least-significant-bits of a subset of samples in a 16 bits-per-sample audio file. Similarly, in \cite{gong2017crafting}, the effectiveness of the proposed attack for speech paralinguistic tasks is measured for different perturbation amounts under the $\ell_\infty$ norm. The restrictions applied in both cases guarantee that the maximum change applicable to each value of the signal is constrained. However, such thresholds are not representative for acoustic signals, as they do not guarantee a low perceptual distortion on audio attacks.

In \cite{carlini2018audio,neekhara2019universal,yang2018characterizing}, in which audio adversarial perturbations for speech recognition models are addressed, the relative loudness of the adversarial perturbation $v$ with respect to the original signal $x$ is measured in Decibels (dB), which is a more representative metric for acoustic signals:
\begin{equation}
\label{eq:db_max}
dB_{x,max}(v)=dB_{max}(v)-dB_{max}(x),
\end{equation}
where
\begin{equation}
dB_{max}(x)=\max_{i} \ 20 \ log_{10}(x_i)
\end{equation}
In \cite{du2019sirenattack}, the signal to noise ratio (SNR) is used to measure the relative distortion of adversarial perturbations for speech recognition models, computed as:
\begin{equation}
\label{eq:SNR2}
SNR(x,v)=10 \ \text{log}_{10}\frac{P(x)}{P(v)},
\end{equation}
where $P(x)$ and $P(v)$ represent the power of the clean signal $x$ and the perturbation $v$, respectively. The SNR has been used in other works on audio adversarial examples \cite{kereliuk2015deep, carlini2016hidden, yuan2018commandersong, yakura2018robust, abdoli2019universal}. However, these works are not based on speech signals, as their approaches rely on data with very different characteristics, such as urban sound classification, music content analysis or the injection of malicious commands into songs. Therefore, the results are not directly comparable to spoken speech recognition, the tasks addressed in this paper.

\subsubsection{Models of human perception and hearing system}
\label{section:assessing_human_perception}

The human hearing system is able to identify sounds in a range from 20Hz to 20kHz, so that perturbations outside this range can not be perceived \cite{rosen2010signals,rossing2007springer}. Based on this fact, in \cite{zhang2017dolphinattack} and \cite{roy2017backdoor}, high frequencies are used to generate audio inaudible to humans but which is captured and classified by a device. Although these attacks may not fit in our specification of adversarial examples (since humans cannot perceive the generated audio, and therefore cannot judge it as benign either), they introduce the idea of using frequency ranges that are out of the human hearing range in adversarial scenarios. 

A different strategy is employed in \cite{schonherr2018adversarial}, where a psychoacoustic model \cite{zwicker2013psychoacoustics} is used to compute the hearing thresholds of different zones of the clean audio signal, which is used to restrict the perturbation to the least perceptible parts.

\subsubsection{Human evaluation}
\label{section:assessing_human_evaluation}

In \cite{cisse2017houdini} and \cite{kreuk2018fooling}, an ABX test is performed, which is a standard method to identify detectable differences between two choices of sensory stimuli. In this method a subject is asked to listen to two audios A and B, and afterwards a third audio X, which will be either A or B, randomly selected. The objective of this test is to assess if the user is able to distinguish between A and B. Optimally, the accuracy ratio would be 50\%, equal to the probability of selecting randomly between the two choices. In our scenario, the two initial audios A and B would correspond to the clean and perturbed audio (in any order).

In \cite{yakura2018robust} and \cite{yuan2018commandersong} the adversarial perturbations are embeded in songs, which can be deployed in the physical world without raising suspicions for humans listeners (e.g., in elevators or TV advertisements) to force a target model to understand speech commands. In both works a human evaluation is carried out on Amazon Mechanical Turk. According to the results presented by the authors, almost none of the participants perceived speech in the perturbed signals. However, a considerable percentage of people reported that an abnormal noise could be noticed in the songs.

In \cite{schonherr2018adversarial}, a \textit{Multiple Stimuli with Hidden Reference and Anchor} (MUSHRA) test is carried out \cite{schinkel2013audio}, to perform a subjective assessment of the audio quality of adversarial examples. The goal of the test is to score the quality of perturbed audio signals (\textit{anchors}, e.g., adversarial examples) with respect to the original signal (\textit{hidden reference}, in this context, the original audio).\footnote{It is worth mentioning that this test is mainly used to assess the intermediate quality level of coding systems, whereas for small impairments, which should be the case of audio adversarial perturbations, dedicated tests have been proposed.} According to the results, the adversarial examples obtained considerably lower scores than the clean audio signals.

Finally, in \cite{vaidya2015cocaine, carlini2016hidden, alzantot2018did, gong2017crafting, du2019sirenattack}, experiments with human subjects are performed with the aim of analyzing their response to the task, in order to assess if the adversarial perturbation has any influence on the responses provided by human listeners. However, no analysis of the perceptual distortion introduced by the perturbations is reported, except in \cite{du2019sirenattack}, in which the subjects are asked to evaluate the noise level of the audio signals.

\subsection{Summary}
Despite the fact that different methods have been proposed to measure the distortion levels introduced by audio adversarial perturbations, we found that the majority of the approaches are not enough to adequately represent the human perception of these attacks. Apart from that, some of the thresholds and acceptable distortion levels assumed in previous works do not always guarantee that the perturbations are imperceptible, and therefore, the detectability of the attacks can be questionable. With this paper, we intend to provide evidence and raise awareness about these gaps.  We hope that the results reported may contribute to establish a more thorough measurement of the distortion, and therefore, to a more realistic study of audio adversarial examples.

\section{Adversarial examples of speech commands}  \label{section:experimental_setup}

Our goal is to evaluate the detectability of audio adversarial perturbation, and to determine  to what extent the metrics commonly used in the literature agree with the human evaluation. To accomplish this goal, we should first establish a number of stepping stones:

\begin{enumerate}
 \item Identify a suitable and representative audio task.
 \item Identify a model appropriate for the task
 \item Collect or identify a dataset to train a model.
 \item Using the model, generate the adversarial examples for the task.
 \item Estimate the actual fooling rate of the adversarial examples. 
\end{enumerate}

\subsection{Selection of the task, model, and dataset}

The task we have selected is \emph{speech command classification} since it is an exemplar machine learning task which is part of the repertoire  of extensively used speech-based virtual assistants.

The DNN model we have selected is based on the architecture proposed for small-footprint keyword recognition \cite{Sainath2015convolutional}. This model takes as input an audio waveform, computes the spectrogram of the signal and extracts a set of MFCC features for different time intervals. This results in a two-dimensional representation of the audio signal, which is fed into the following topology: two convolutional layers with a ReLU activation function, a fully-connected layer and a softmax layer. The same architecture has been used in related works on adversarial examples \cite{alzantot2018did,du2019sirenattack} and as a baseline model in other research tasks \cite{warden2018speech,zhang2017hello}. 

We used the Speech Command Dataset \cite{warden2018speech}, which is a widely used dataset in the study of adversarial attacks for speech recognition systems \cite{alzantot2018did,yang2018characterizing,du2019sirenattack}.
The dataset is composed of recordings of 30 different spoken commands, provided by a large number of different people. Audio files are stored in a 16-bit WAV file, with a sample-rate of 16kHz and a fixed duration of one second. As in previous publications \cite{warden2018speech,alzantot2018did}, we selected the following subset of commands to develop our work: ``Yes'', ``No'', ``Up'', ``Down'', ``Left'', ``Right'', ``On'', ``Off'', ``Stop'', and ``Go''. Additionally, We will also consider two special classes: ``Unknown'' (a spoken command not considered in the previous set), and ``Silence'' (no speech detected in the audio). Note that this selection comprises a wide variety of commands in terms of phonetic similarity.

\subsection{Generating  \textit{single-class} universal perturbations}

As previously mentioned, we focus on \textit{single-class} universal perturbations, an attack approach that attempts to generate a single perturbation which is able to fool the model for any input corresponding to a particular class $y$. We decided to focus on universal perturbations because an initial experimentation with individual perturbations (crafted using Deepfool algorithm) led us to the conclusion that the perturbations were undoubtedly imperceptible. This conclusion has been reported before in the literature \cite{fezza2019perceptual} for the case of image adversarial examples. Therefore, we selected the more challenging task of generating  universal perturbations, which requires higher distortion levels. Moreover, we selected \textit{single-class} universal attacks in order to study in more detail the results on different commands. The particular choice of the class to which the target perturbation is applied is a factor that may influence the perceptual distortion of the perturbations. 

The selected attack method is based on the strategy proposed in \cite{moosavi2017universal}, a state-of-the-art method to generate universal perturbations based on accumulating individual perturbations created for a set of \emph{training} samples using the Deepfool algorithm \cite{moosavi2016deepfool}. We use the \emph{UAP-HC} reformulation of this strategy for audio samples as presented in \cite{Vadillo_and_Santana:2019}, where more details about the process to generate the perturbations can be found.

We generated 5 different universal perturbations per class, starting from a different \emph{training} set of 1000 samples in each case. During the crafting process, the universal perturbations were bounded by the $\ell_2$ norm, with a threshold value of $\xi=0.1$. In addition, the Deepfool algorithm was limited to a maximum number of 100 iterations. The overshoot parameter of the Deepfool algorithm was set to 0.1. Finally, the \emph{UAP-HC} algorithm was restricted to 5 epochs, that is, 5 complete passes through the entire training set.

\subsection{Effectiveness of the perturbations fooling the model}

To measure the effectiveness of the universal perturbations, we compute the percentage of audios for which the prediction changes when the perturbation is applied. We will refer to this metric as fooling ratio (FR) \cite{moosavi2017universal}. 
The effectiveness of the generated perturbations is shown in Table \ref{tab:level1_results}, for the training set (the set of samples used to optimize the universal perturbation) and for the validation set (the set of samples used to compute the effectiveness of the attack for inputs not used during the optimization process).\footnote{The samples used to optimize the perturbation will be selected from the training set used to train the DNN. Equivalently, the \textit{validation} set of the algorithm will be selected from the validation set used during the training process of the model.} Results are shown for the average effectiveness of the 5 perturbations generated for each class, as well as for the one that maximizes the FR on the training set. 

According to the results, the generated adversarial examples are highly effective for the majority of the classes, with a maximum FR above 70\% for 7 out of 12 classes in both training and validation sets. Note that we obtain a considerably high effectiveness also in the class \textit{unknown}, which is composed of a diverse set of spoken commands. However, the hardest class to fool is \textit{silence}, in which the maximum FR is below 25\% in both training and validation sets. This may be due to the fact that, according to the nature of the audios corresponding to that class, trying to fool the model by adding a small amount of \textit{noise} is a challenging task.

\begin{table}[t]
\centering
\caption{Effectiveness of the generated single-class universal perturbations.}
\label{tab:level1_results}
\begin{tabular}{lcccc}
\toprule
\multirow{2}[2]{*}{Class} & \multicolumn{2}{c}{Max. FR\%} & \multicolumn{2}{c}{Mean FR\%} \\ 
\cmidrule(lr){2-3} \cmidrule(lr){4-5} 
 & Train & Valid & Train & Valid \\
 \midrule
Silence & 23.80 & 19.46 & 22.24 & 19.61 \\
Unknown & 72.70 & 73.06 & 70.58 & 73.51 \\
Yes & 74.50 & 74.36 & 68.26 & 66.40 \\ 
No & 86.50 & 83.77 & 81.48 & 79.40 \\ 
Up & 84.20 & 75.45 & 82.20 & 74.73 \\ 
Down & 71.50 & 65.55 & 68.06 & 64.51 \\ 
Left & 52.30 & 49.73 & 42.20 & 40.59 \\ 
Right & 68.70 & 63.82 & 60.62 & 56.47 \\
On & 76.00 & 75.65 & 54.42 & 53.28 \\ 
Off & 80.10 & 73.48 & 75.18 & 70.85 \\ 
Stop & 61.40 & 61.82 & 56.92 & 57.30 \\
Go & 87.80 & 80.06 & 86.24 & 80.90 \\ 
\bottomrule
\end{tabular}
\end{table}

\begin{figure}[]
\centering
\includegraphics[scale=0.52]{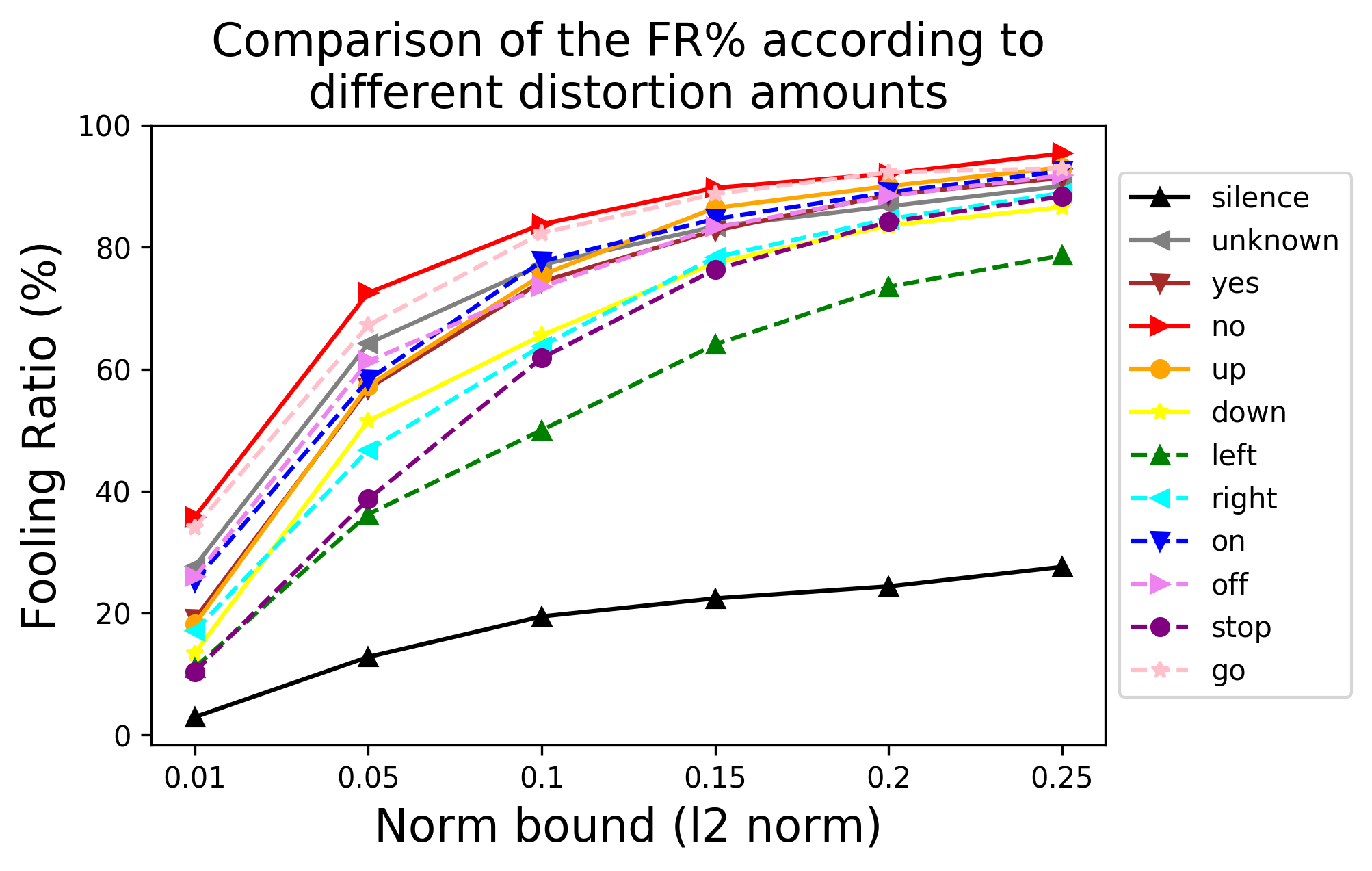}
\includegraphics[scale=0.52]{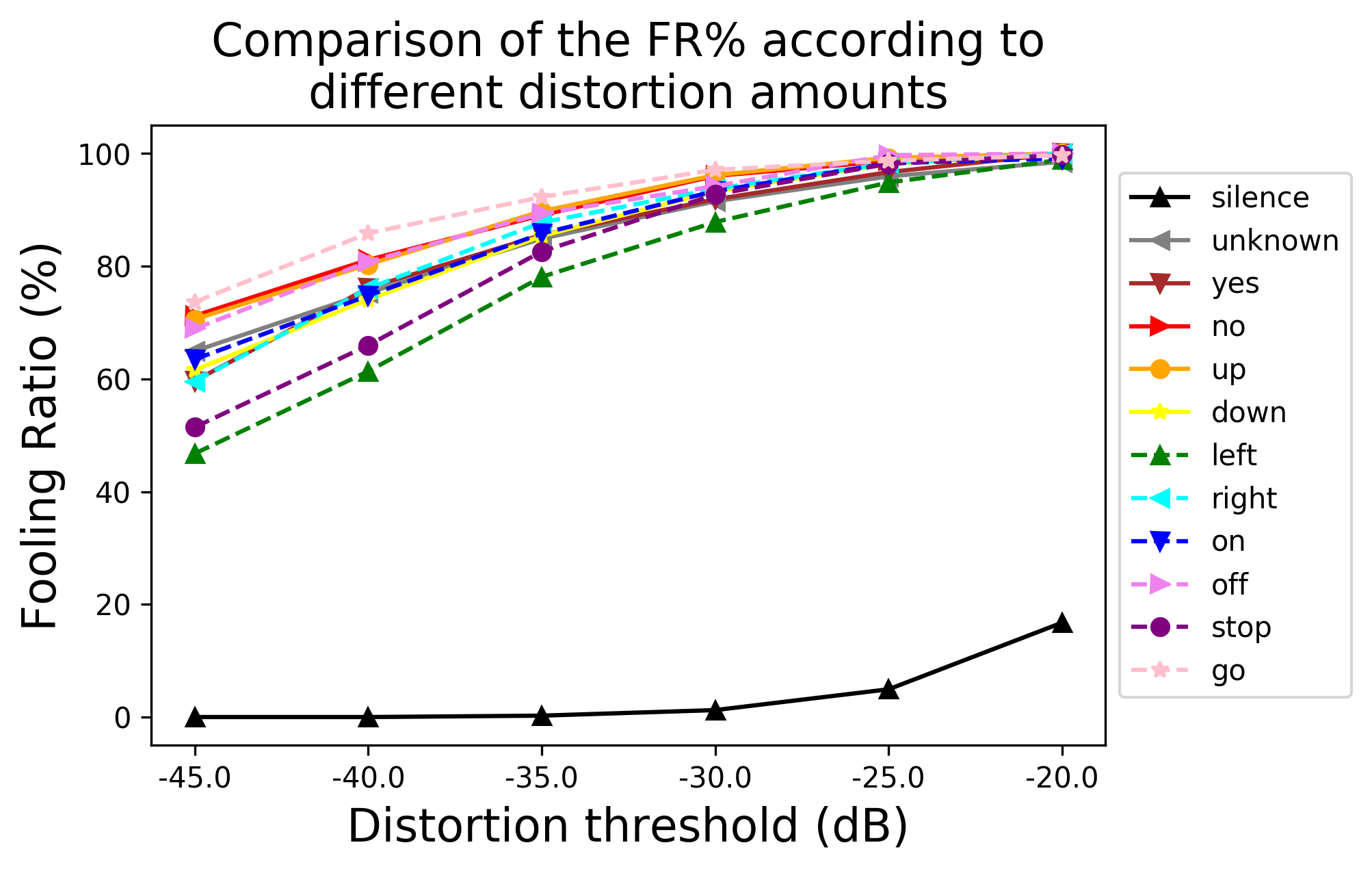}
\caption{Variation in the effectiveness (FR\%) in the validation set of the generated \textit{single-class} universal adversarial perturbations according to two different criteria: $\ell_2$ norm of the perturbation (top) and $dB_{x,max}(v)$ metric with respect to each input signal $x$ (bottom).}
\label{fig:effectiveness_fr_evolutions}
\end{figure}

It is important to bear in mind that the effectiveness of a universal perturbation is directly correlated to the distortion amount introduced. We show in Fig. \ref{fig:effectiveness_fr_evolutions}, for each class, the way in which the FR increases as the distortion amount introduced by the perturbations increases. These results have been obtained by scaling the magnitude of a universal perturbation $v$ according to two distortion criteria: the $\ell_2$ norm of the perturbation and the decibel difference between the perturbation and each sample of the dataset. In the first case, the perturbation signal is scaled in order to ensure that its norm equals the desired threshold, and it is equally applied to every input sample. In the second case, the perturbation signal is scaled for every input sample $x$, in order to ensure that the $dB_{x,max}(v)$ metric equals the specified threshold.

The fact that the FR is directly correlated with the distortion level implies that there is a trade-off between the effectiveness and the detectability of the attacks. Therefore, to adequately study the risk posed by audio adversarial attacks, it is important to establish realistic and rigorous criteria for assessing the human perception of such attacks.

\section{Evaluation of the distortion using similarity metrics}
\label{section:evaluation_metrics}

While the ability to fool the model is an essential ingredient of adversarial examples, the other requirement is that the perturbation is not noticed by humans. In this section, we evaluate the distortion produced by the generated adversarial perturbations, according to different criteria.

\subsection{Evaluating the distortion: the standard, uninformed way}

We first computed the distortion according to the standard approaches employed in previous works on adversarial examples in speech related tasks \cite{carlini2018audio,neekhara2019universal,yang2018characterizing}, as described in equation \eqref{eq:db_max} (Section \ref{section:previous_works_audio}). In \cite{carlini2018audio}, where individual adversarial perturbations are created for speech transcription scenarios, the mean distortion of the generated perturbations is -31dB, and the 95\% interval for distortion ranges from -15dB to -45dB.\footnote{Note that according to this metric, the lower the distortion value, the less detectable the perturbation.} The same range of distortion is reported in \cite{yang2018characterizing}. In \cite{neekhara2019universal}, where universal adversarial perturbations are generated also for speech transcription models, the distortion level of the perturbations is bounded under different thresholds, obtaining a mean distortion of $\approx -42$dB in the best case, and a mean distortion of $\approx -30$dB in the worst case. Overall, distortion levels below -32dB are considered acceptable in these works.

\begin{figure}[!h]
\centering
\includegraphics[scale=0.5]{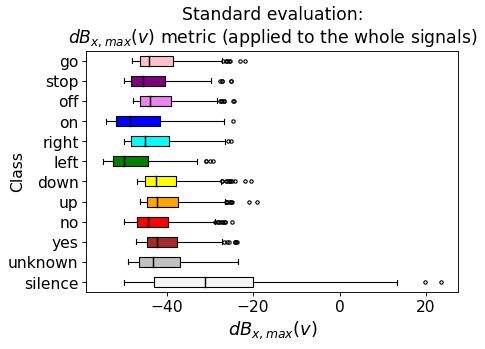}
\caption[Distortion level of the generated \textit{single-class} universal perturbations, evaluated in the validation set.]{Distortion level of the generated \textit{single-class} universal perturbations, evaluated in the validation set using the standard evaluation approach: $dB_{x,max}(v)$ applied to the whole signals. Results are averaged for the 5 perturbations generated for each class.}
\label{fig:universal1_standard_distortion}
\end{figure}

Fig. \ref{fig:universal1_standard_distortion} shows the distortion level of the generated perturbation with respect to each input sample in the validation set, according to the same approach. Results are computed independently for each class, and averaged for the 5 trials carried out in each of them. Table \ref{tab:universal1_standard_distortion} shows the mean distortion level obtained for each class. As can be seen, the mean distortion is below -40dB in all the classes except \textit{silence}, in which the mean distortion is of -29.52dB\footnote{This effect can be explained by the fact that, due to the nature of the samples corresponding to the class \textit{silence}, their loudness level is lower than for the rest of classes.}. Moreover, without considering the class \textit{silence}, more than 90\% of the samples are below -32dB in all the cases. Therefore, our perturbations can be considered as highly acceptable according to this standard.

\begin{table}[]
\centering
\caption{Distortion levels produced by the generated single-class universal perturbations (standard evaluation). Results are averaged for the 5 experiments carried out for each class.}
\label{tab:universal1_standard_distortion}
\begin{tabular}{lcc}
\toprule
Class & 
\begin{tabular}[c]{@{}c@{}}Mean  \\ $dB_{x,max}(v)$ \end{tabular} &
\begin{tabular}[c]{@{}c@{}}\% of samples \\ below -32dB\end{tabular}\\ 
\midrule
Silence & -29.52 & 48.04 \\ 
Unknown & -41.35 & 90.20\\ 
Yes     & -40.58 & 90.45\\ 
No      & -42.56 & 93.09\\ 
Up      & -40.24 & 89.18  \\ 
Down    & -40.63 & 90.64\\ 
Left    & -48.10 & 99.03 \\ 
Right   & -43.30 & 95.20 \\ 
On      & -46.31 & 96.21 \\ 
Off     & -42.01 & 94.03 \\ 
Stop    & -43.92 & 96.11 \\ 
Go      & -41.88 & 93.28 \\ 
\bottomrule
\end{tabular}
\end{table}

\subsection{Evaluating the distortion: detailed and signal-part-informed way}

 In order to measure the distortion in more detail, we employed the approach presented in \cite{Vadillo_and_Santana:2019}. In this case, the distortion induced by the perturbation $v$ in the original sample $x$ is computed in terms of the difference between both the maximum (as defined in equation \eqref{eq:db_max}) and the mean decibel values, defined as:

\begin{equation}
 \label{eq:mean_metric}
 dB_{x,mean}(v) = dB_{mean}(v)-dB_{mean}(x),    
\end{equation}
where
\begin{equation}
 \label{eq:db_mean}
dB_{mean}(x)=20\cdot \text{log}\left(\frac{1}{d}\sum_{i=1}^d{|x_i|}\right)
\end{equation}

Furthermore, previous work on evaluating the naturalness of adversarial examples in the audio domain compute the distortion between two signals by applying the metrics to the entire signals \cite{carlini2018audio,yang2018characterizing, neekhara2019universal}. In this paper, we advocate the application of both metrics in two different parts of each audio signal: the \emph{vocal} part and the \emph{background} part. This differentiation is due to the fact that, for spoken commands, the amount of sound outside the vocal part is considerably lower. Thus, the same amount of perturbation would be perceived differently depending on the infected part. By mapping the distortive effect of the perturbations to these parts of the signals we also get a better assessment of how the attack works better.

As we are handling short single-command audio signals, the vocal part of an audio signal $x=\left\lbrace x_1...x_d \right\rbrace$ will be delimited by the continuous range $[x_a,x_b]$ containing 95\% of the energy of the signal, that is:
\begin{equation}
\frac{\sum_{i=a}^{b}x_i^2}{\sum_{i=1}^{d}x_i^2}\approx 0.95.  
\end{equation}
Thus, we will assume that ranges $[x_1,x_a)\cup (x_b,x_d]$ will be composed just of background noise. Notice that this partition is well suited for single command audios in which it is assumed that the vocal part of the signal is contiguous. Audio signals belonging to the \textit{silence} class will be omitted from the analysis of the vocal part, as they are composed only of background noise, without any vocal part.

\subsection{Results of the different signal-part approach}

\begin{figure}[]
\centering
\includegraphics[scale=0.43]{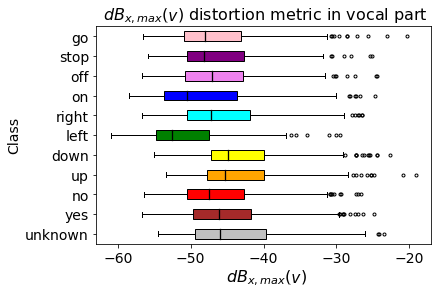}
\includegraphics[scale=0.43]{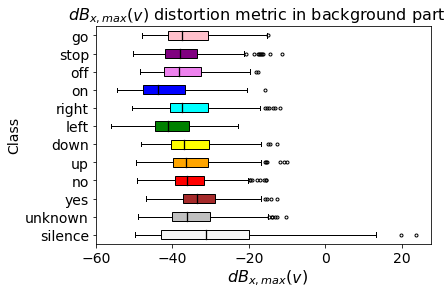}
\includegraphics[scale=0.43]{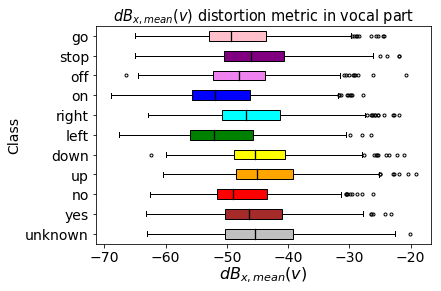}
\includegraphics[scale=0.43]{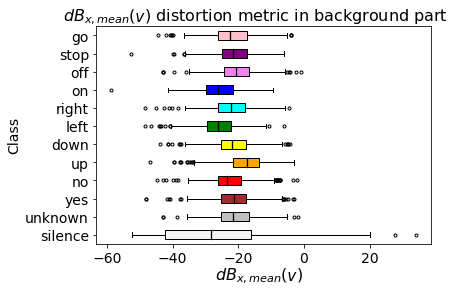}
\caption[Distortion level of the generated single-class 1 universal perturbations, evaluated in the validation set.]{Distortion level of the generated \textit{single-class} universal perturbations, evaluated in the validation set using $dB_{x,max}(v)$ metric (top row) and $dB_{x,mean}(v)$ metric (bottom row). For each audio, the distortion has been measured in the vocal part as well as in the background part. Results are averaged for the 5 perturbations generated for each class.}
\label{fig:universal1_distortion}
\end{figure}

The results obtained with the described evaluation approach are shown in Fig.~\ref{fig:universal1_distortion}. The first row of the figure shows the results obtained using $dB_{x,max}(v)$ metric, and the bottom row the results obtained using $dB_{x,mean}(v)$ metric. Notice the difference between the horizontal axis scales of the figures.

By comparing the perturbations in the vocal part and the background part, it can be seen that perturbations in the vocal part are less noticeable, with a decibel difference significantly lower, which occurs using both $dB_{max}$ and $dB_{mean}$ distortion metrics.

Regarding the distortion amount in the vocal part, the obtained results are significantly below the threshold of $-32$dB in almost all the samples, independently of the metric. Compared to the sound intensity level of a normal conversation, a distortion of $-30$dB corresponds to the weakest audible signal between 10kHz and 100Hz frequency range \cite{smith1997scientist}, which is roughly the difference between the ambient noise in a quiet room and a person talking \cite{carlini2018audio}.

While the distortion level outside the vocal part is still acceptable under the $dB_{max}$ metric, according to the $dB_{mean}$ metric the distortion exceeds the threshold of -32dB for a great majority of the samples. In fact, in about half of the cases the difference in decibels is greater than -20dB, which may indicate that the perturbations could be highly detectable in those parts.

\section{Human evaluation of voice command adversarial examples}
\label{section:human_experiment_distortion}

 While the methods presented in the previous section provide a more accurate and detailed assessment on the quality of the adversarial examples, the metrics used are not expected to capture all the subtleties of a proper human evaluation. Therefore, we designed an experiment in which human subjects listen to audio adversarial examples and judge them according to different criteria.  The main goal of the experiment was to study to which extent the perturbations are detectable by humans. In this section we describe the experimental design and its results.
 
\subsection{Experimental design}

  A set of eighteen subjects (12 men and 6 women), independent of the research, was selected to conduct the experiment. Each participant was instructed to listen to different audio clips and answer some questions about them. The experiment is composed of two parts:

\begin{itemize}
\item In the first part, the naturalness of the generated universal adversarial examples is investigated. The other question investigated is to what extent the distortion produced by the perturbation affects the understandability of the spoken commands. To address these questions, the participants are asked to listen a set of 12 audio clips, either clean or adversarially perturbed, and provide the following information: 

\begin{itemize}
\item Identify the command that can be heard in the audio clip, in order to determine if the adversarial perturbations affect the understandability of the spoken commands.

\item Assess the level of naturalness of the audio clip, in order to study whether the adversarial examples are perceived as perturbed audios in comparison to clean instances. As both clean and perturbed audios will be tested, the comparison between the results obtained in both cases may reflect if the perturbations are perceived just as a regular background noise or other ordinary perturbations, or whether they are perceived as artificial or malicious. In the experiment, the subjects evaluated the naturalness on a scale from 1 to 5, with the following scale provided as reference:

\begin{itemize}
\item[] 1) Clearly perturbed audio with an artificial sound or noise.
\item[] 2) The audio is slightly perturbed by an artificial sound or noise, not likely to be caused by the low quality of the microphones or ambient sounds.
\item[] 3) Not sure
\item[] 4) No obvious signs of an artificial perturbation. The detectable perturbations are likely to be caused by a low- or mid-quality microphone, ambient sounds or ordinary noises. 
\item[] 5) The audio clip clearly does not contain any artificial perturbation.
\end{itemize}

\end{itemize}

\item In the second part of the experiment, each participant performed an ABX test, a method to identify detectable differences between two choices of sensory stimuli. In this method, a subject is asked to listen to two audios A and B, and afterwards a third audio X, which will be either A or B, randomly selected. The goal of the test is to evaluate if the subject is able to determine if X corresponds to A or to B. In our experiment, the two initial audios A and B will correspond to the clean and perturbed audio, in any order. Thus, this test will determine if the perturbations are detectable in comparison to the clean audio sample. Six trials were carried out in each experiment, that is, six sets of three audio clips A, B and X.
\end{itemize}

Due to the fact that the audio clips of the dataset contain different characteristics, such as the intensity of the spoken command or the amount of background noise, the perception of a perturbation may change according to these features. For this reason, we decided to classify the audios considering three levels of intensity: low, medium and high. The $dB_{mean}$ distortion metric presented in \eqref{eq:db_mean} will be used to measure the mean distortion of the audio signals. According to this metric, 99\% of the intensities of the audio samples lie approximately in the decibel range $[30,85]$. By performing a rough uniform binning of the intensity range (known as equal-width binning in the literature \cite{dougherty1995supervised}), the levels were defined as follows:  
\begin{itemize}
\item Low intensity level: audios with a mean distortion below 50dB.
\item Medium intensity level: audios with a mean distortion between 50dB and 70dB.
\item High intensity level: audios with a mean distortion above 70dB.
\end{itemize}
To ensure a uniform representation of the different levels of intensity, each experiment was composed of audio signals of only one of these levels. Nine different experiments were created, (three experiments per intensity level), and each of them was assigned to two different participants, making a total of 18 experiments and participants. A summary of the final experimental setup is provided in Table \ref{tab:human_experiment_design}. The minor unbalance in the number of original and modified audios in the first part of the experiment is assumed to ensure greater uniformity in the frequency of each command, depending on the different factors influencing the experiment, as shown in Table \ref{tab:exp_part1_command_frequency},  as well as to ensure that the model correctly classified the original audio samples but incorrectly classified the adversarial examples.

\begin{table}[!b]
\centering
\caption{Summary of the experimental setup designed for the human evaluation of the distortion produced by the universal perturbations.}
\label{tab:human_experiment_design}
\begin{tabular}{cccccc}
\toprule
\multirow{2}[2]{*}{Experiment} & \multirow{2}[2]{*}{Intensity} & 
\multicolumn{3}{c}{Audio samples (part 1)} & \multirow{2}[2]{*}{\begin{tabular}[c]{@{}c@{}}ABX trials \\ (part 2)\end{tabular}} \\ 
\cmidrule(lr){3-5}
 &  & Clean & Adv. & Total &  \\ 
\midrule
1 & Low & 6 & 6 & 12 & 6 \\ 
2 & Low & 6 & 6 & 12 & 6 \\ 
3 & Low & 5 & 7 & 12 & 6 \\ 
4 & Medium & 6 & 6 & 12 & 6 \\ 
5 & Medium & 6 & 6 & 12 & 6 \\ 
6 & Medium & 7 & 5 & 12 & 6 \\ 
7 & High & 6 & 6 & 12 & 6 \\ 
8 & High & 6 & 6 & 12 & 6 \\ 
9 & High & 7 & 5 & 12 & 6 \\ 
\toprule
\end{tabular}
\end{table}

\begin{table*}[t]
\centering
\caption{Number of audios per command used in the experiment (part 1).}
\label{tab:exp_part1_command_frequency}
\scalebox{0.94}{
\begin{tabular}{lcccccccccccc}
\toprule
Type & Sil. & Unk. & Yes & No & Up & Down & Left & Right & On & Off & Stop & Go \\ 
\midrule
Low intensity & 6 & 6 & 6 & 6 & 6 & 6 & 6 & 4 & 6 & 6 & 8 & 6 \\
Medium intensity & 6 & 4 & 8 & 6 & 6 & 6 & 6 & 6 & 6 & 6 & 6 & 6 \\ 
High intensity & 6 & 12 & 8 & 6 & 0 & 12 & 4 & 8 & 2 & 2 & 4 & 8 \\ 
\addlinespace
Clean & 12 & 2 & 8 & 12 & 4 & 22 & 12 & 4 & 4 & 12 & 14 & 4 \\ 
Adversarial & 6 & 20 & 14 & 6 & 8 & 2 & 4 & 14 & 10 & 2 & 4 & 16 \\ 
\addlinespace
Total Frequency & 18 & 22 & 22 & 18 & 12 & 24 & 16 & 18 & 14 & 14 & 18 & 20 \\ 
\bottomrule
\end{tabular}
}
\end{table*}

\subsection{Analysis of the results}

\subsubsection{Command classification task}

The first factor to be analyzed is the accuracy percentage obtained by humans in the command classification task (first part of the experiment), that is, which percentage of samples have been correctly labeled by humans. 

\begin{figure}[h]
\centering
\includegraphics[scale=0.43]{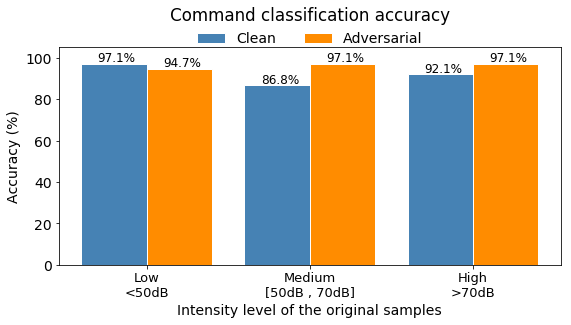}
\caption[Accuracy percentages achieved by the participants of the experiment in the speech command classification task.]{Accuracy percentages achieved by the participants of the experiment in the speech command classification task. Results have been split for each sample type (clean or adversarial) as well as for the intensity levels of the original audios in the experiments (low, medium or high).}
\label{fig:audibilty_acc_classification}
\end{figure}

According to the results, the total number of instances wrongly classified considering all the instances, clean and adversarial, is 13 out of 216 (9 corresponding to clean samples and 4 to adversarial samples), which corresponds to a total accuracy in the command classification of $\sim 94$\%. In order to provide more detailed information, the accuracies obtained for each intensity level are shown in Fig. \ref{fig:audibilty_acc_classification}, in which the percentages are computed independently for clean instances and for adversarial examples.  Overall, these results indicate that the adversarially perturbed spoken commands are clearly recognizable and well classified by humans, independently of the intensity level of the original audio. In other words, although the adversarial perturbations are able to fool the target model, they do not affect the human understanding of the command.  The obtained results are consistent with those achieved in \cite{du2019sirenattack}, where the success rate of a set of people in classifying audio commands is reported using the same dataset as us, but without considering \textit{silence} or \textit{unknown} as classes and without differentiating between the intensity level of the original signals. According to the results reported in \cite{du2019sirenattack}, the accuracy in recognizing the commands was 93.5\% for clean samples and 92.0\% for adversarial examples.

\subsubsection{Naturalness}

\begin{figure}[h]
\centering
\includegraphics[scale=0.43]{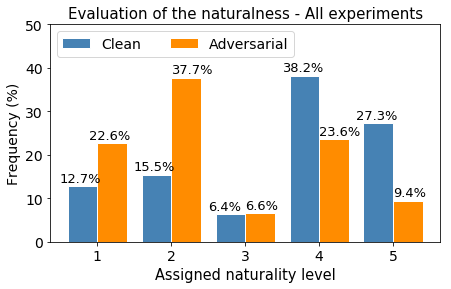}
\includegraphics[scale=0.43]{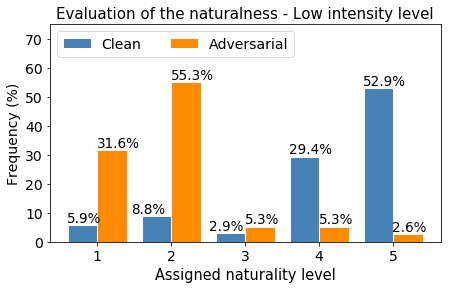}
\includegraphics[scale=0.43]{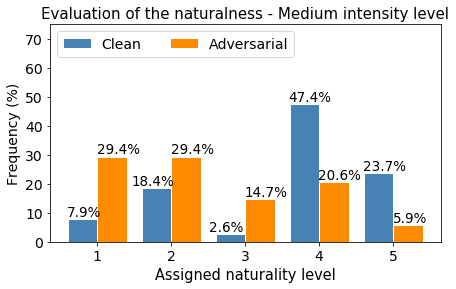}
\includegraphics[scale=0.43]{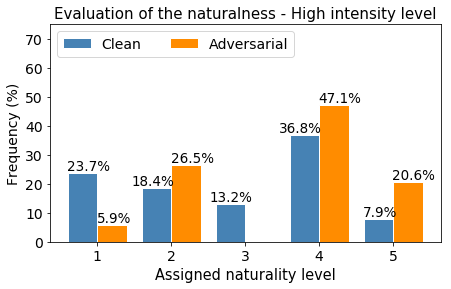}
\caption[Analysis of the naturalness level assigned to the audio samples of the speech command classification task.]{Analysis of the naturalness level assigned to the audio samples of the speech command classification task in all the experiments, split by sample type (clean or adversarial). The results are computed for all the experiments (top left) as well as for each intensity level individually: low (top right), medium (bottom left) and high (bottom right).}
\label{fig:audibilty_naturalness_types}
\end{figure}

Furthermore, the results obtained in the analysis of the naturalness level assigned to the instances is displayed in Fig. \ref{fig:audibilty_naturalness_types}. The figure shows the frequencies with which samples are classified in each naturalness level, split according to the sample type (clean or adversarial). In addition, the results are jointly computed for all the experiments (top left) as well as for each intensity level individually: low (top right), medium (bottom left) and high (bottom right). Considering all the experiments, it can be observed that the adversarial examples obtained lower scores in comparison to the clean samples. We verified by an exact multinomial statistical test that there exist significant differences regarding the scores assigned to clean and adversarial audios (achieving a p-value below a tolerance of $0.01$). Indeed, while $65.5$\% of the clean samples are classified with a naturalness level of 4 or 5, only $33$\% of adversarial examples have been classified in the same range. These results indicate that, in general, the adversarial perturbations are perceived in the audio signals as artificial sounds or noises with a considerably higher frequency than clean samples.

Doing the same analysis independently for each intensity level, it can be observed that the main difference is given in the lowest intensity level, in which $86.9$\% of the adversarial examples achieved a score of $1$ or $2$, while only $14.7$\% of clean samples were classified in that range. For the highest intensity level, however, the percentage of adversarial examples scored with a $4$ or $5$ is even greater than the corresponding percentage for clean samples. Thus, the human perception of the adversarial examples is clearly related to the intensity level of the original audio signals. This is a remarkable fact that should be taken into consideration in the evaluation of audio adversarial examples.

\subsubsection{ABX test}

\begin{figure}[h]
\centering
\includegraphics[scale=0.45]{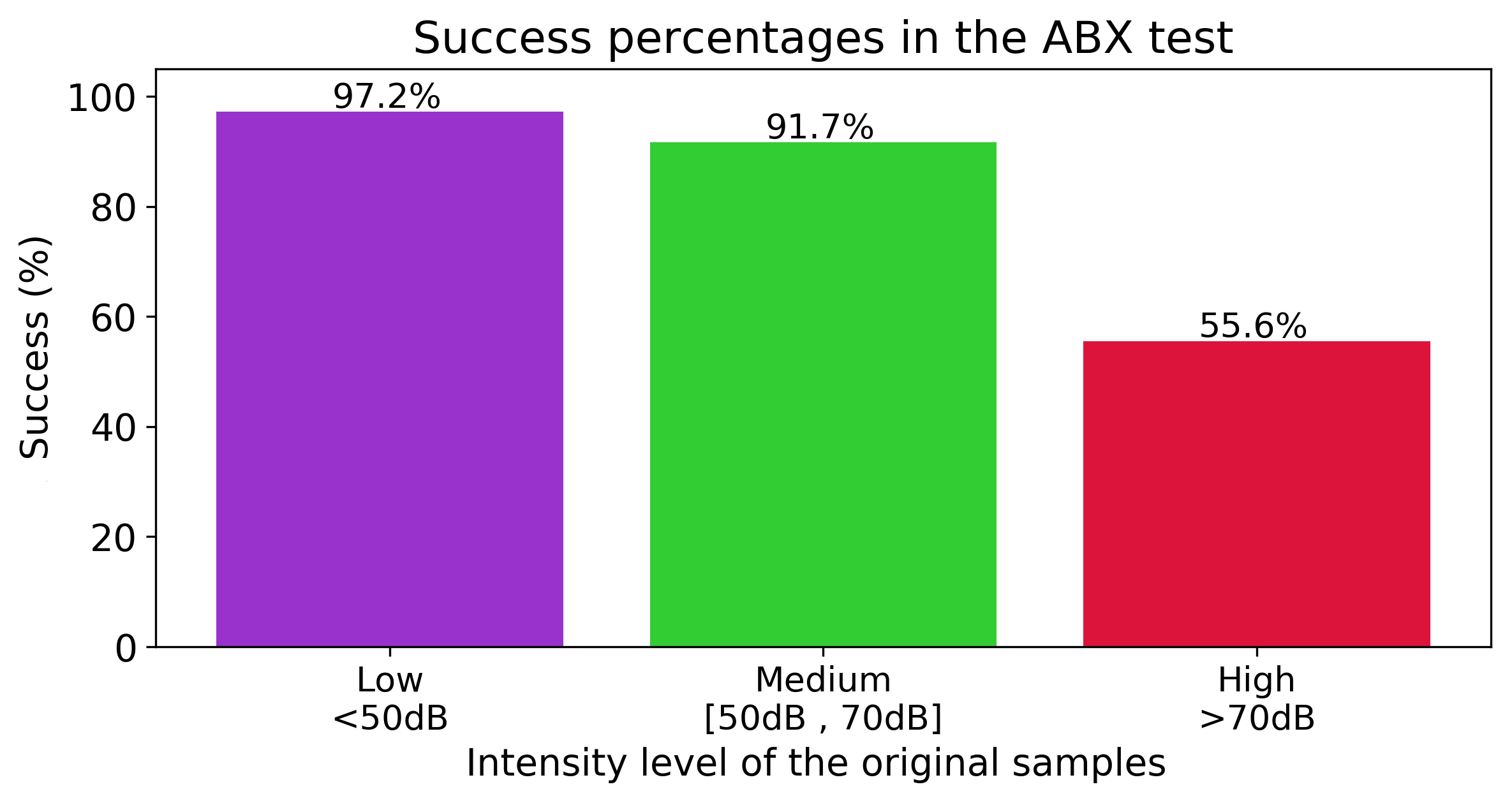}
\includegraphics[scale=0.45]{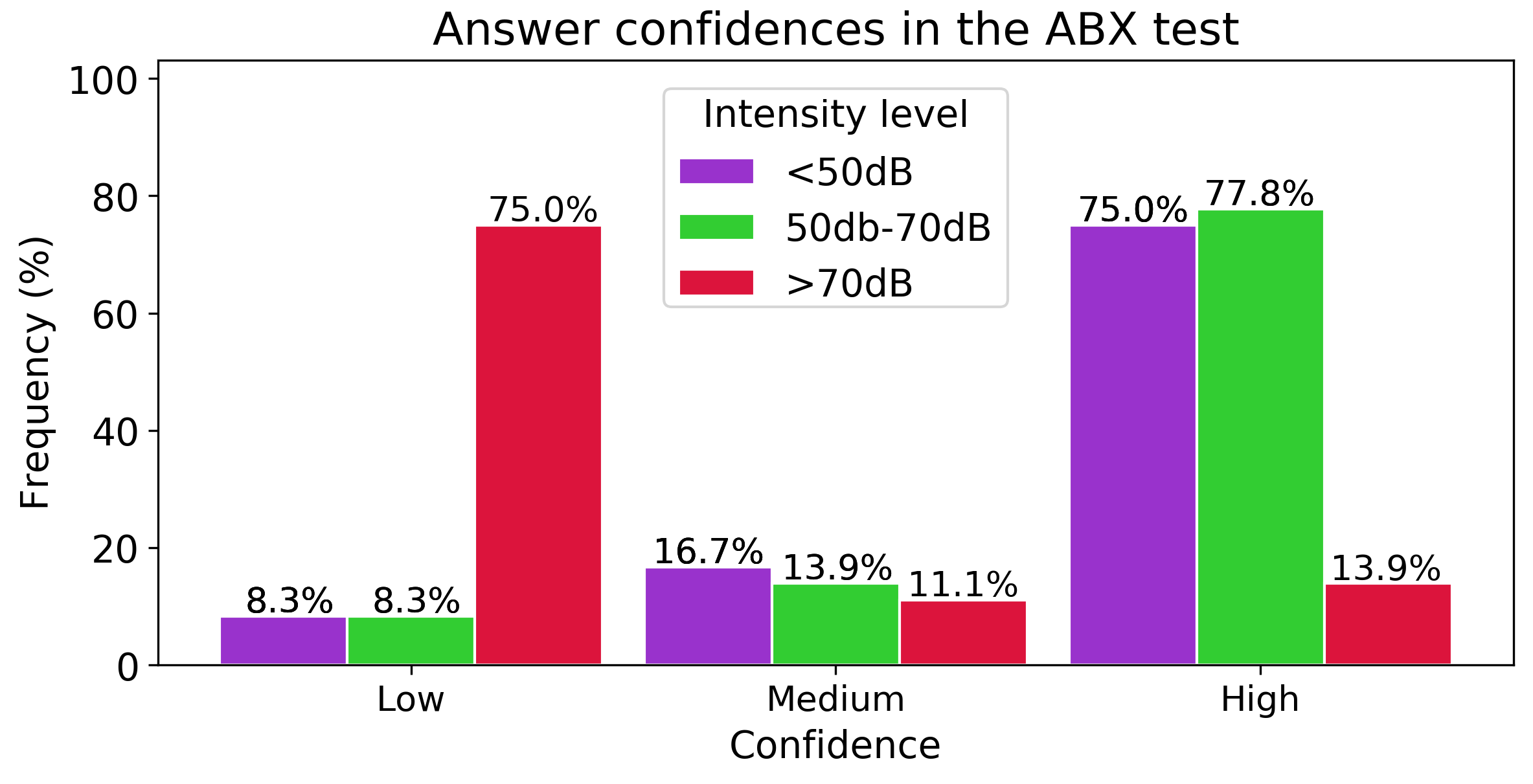}
\caption[Success percentages obtained in the ABX test (top) and confidence levels of the answers in the test (bottom), both computed independently for each intensity level.]{Success percentages obtained in the ABX test (top) and confidence levels of the answers in the test (bottom), both computed independently for each intensity level.}
\label{fig:audibility_abx}
\end{figure}

In order to better evaluate if the perturbations are perceivable, the results obtained in the ABX test (second part of the experiment) have been analyzed. This is summarized in Fig. \ref{fig:audibility_abx}. The first row of the figure shows the percentage of success cases in the ABX test, that is, the percentage of cases in which the unknown audio (audio X) has been correctly classified. The second row shows the confidence level of the answers. All these results have been computed independently for each intensity level.

The success rate of the experiments with low and medium intensity levels is of $97.2$\% and $91.7$\% respectively, revealing that the perturbations are clearly perceivable in such cases. On the contrary, only a $55.6$\% success rate is achieved for high intensity levels, close to the optimum value of $50$\%, which is equivalent to a random guessing. We verified by an exact binomial test\footnote{The alternative hypothesis of the test is that the empirical success ratio is greater than $p=0.5$. The same test with the alternative hypothesis that the empirical ratio is not equal to $0.5$ obtained a p-value of $\approx 0.62$.} that the achieved success ratio is no significantly greater (achieving a p-value of $\approx0.31$) than the probability $p=0.5$ corresponding to a binomial distribution $X\sim B(n=36,p=0.5)$, where $n$ is the sample size. This fact indicates that, in such cases, the adversarial examples are not distinguishable from their corresponding clean audio examples. It is worth noting that, given our experimental setup, 95\% (Clopper–Pearson) confidence intervals of the success ratio is $[0.85,0.99]$ for low intensity audios, $[0.78, 0.98]$ for medium intensity audios and $[0.38,0.72]$ for high intensity audios. The results provided can, therefore, be considered representative of the human perception of the distortion.

Consistently with the success rates, the subjects were highly confident in providing their answers in more than $75$\% of the cases in the experiments containing audios with low and medium intensity levels. Contrarily, in $75$\% of the answers the participants reported a low confidence in the experiments containing audios with high intensity levels.

Overall, these analyses demonstrate that the detectability of the perturbations largely depends on the intensity level of the clean audio, being detectable for audios with low and medium intensity levels, but not perceivable for audios with a high intensity level.

It is worth mentioning that, according to the standard approach used in previous related works to measure the detectability of audio adversarial examples, the crafted perturbations were far below the maximum acceptable distortion. However, the results obtained in this section reinforce our proposal about the need to employ more rigorous approaches in order to measure and set a threshold on the distortion produced by the adversarial perturbations in a more representative way. We encourage the reader to listen to some adversarial examples, to empirically assess the perceptual distortion of adversarial perturbations according to different characteristics.\footnote{\url{https://vadel.github.io/adversarialDistortion/AdversarialPerturbations.html}}

\section{Conclusions} \label{section:conclusions}

In this paper we have addressed the measurement of the perceptual distortion of audio adversarial examples, which remains a challenging task despite being a fundamental condition for effective adversarial attacks. For this purpose, we have performed an analysis of the human perception of audio adversarial perturbations for speech command classification tasks, and this analysis has been used to study whether the distortion metrics employed in the literature correlate with the human judgment. 

We have found out that, while the distortion levels of our perturbations are acceptable according to the standard evaluation approaches employed by convention, the same perturbations were highly detectable and judged as artificial by human subjects. For this reason, we have proposed a novel framework to measure the distortion in a more comprehensive way, based on a differential analysis in the vocal and background parts of the audio signals, which provide a more realistic and rigorous evaluation of the perceptual distortion. Our experiments with \textit{single-class} universal perturbations for a set of varied commands also demonstrate that there exist differences regarding the effectiveness of the attacks, related to the relative distortion, and how the perceptual distortion of the perturbations changes depending on the intensity levels of the audio signal in which it is injected.

These results highlight the lack of audio metrics capable of modeling the human perception in a realistic and representative way, and stress the need to include human evaluation as a necessary step for validating methods used to generate adversarial perturbation in the audio domain. We hope that future works could advance in this direction in order to fairly evaluate the risk that adversarial examples suppose.

\section*{Acknowledgments}

The authors would like to thank to the Intelligent Systems Group (University of the Basque Country, Spain) for providing the computational resources needed to develop the project. This work has received support form the predoctoral grant that Jon Vadillo holds by the Basque Government (reference PRE\_2019\_1\_0128). Roberto Santana acknowledges support by the Basque Government (IT1244-19 and ELKARTEK programs), and  Spanish Ministry of Economy and Competitiveness MINECO (project TIN2016-78365-R).

\medskip

\small

\bibliographystyle{unsrt}
\bibliography{references}

\end{document}